\documentclass[lettersize,journal]{IEEEtran}
\IEEEoverridecommandlockouts
\usepackage{cite}
\usepackage{amsmath,amssymb,amsfonts}
\usepackage[ruled,linesnumbered,lined]{algorithm2e}
\usepackage{balance}
\usepackage{graphicx}
\usepackage{subfigure}
\usepackage{textcomp}
\usepackage{xcolor}
\usepackage{amsthm}
\usepackage{bm}
\usepackage{makecell}
\usepackage{color}
\def\BibTeX{{\rm B\kern-.05em{\sc i\kern-.025em b}\kern-.08em
    T\kern-.1667em\lower.7ex\hbox{E}\kern-.125emX}}

\begin{document}
\title{RS-BNN: A Deep Learning Framework for the Optimal Beamforming Design of Rate-Splitting Multiple Access}
\author{Yiwen Wang,~\IEEEmembership{Student Member,~IEEE}, Yijie Mao,~\IEEEmembership{Member,~IEEE,} Sijie Ji,~\IEEEmembership{Member,~IEEE}
\thanks{Copyright (c) 20xx IEEE. Personal use of this material is permitted. However, permission to use this material for any other purposes must be obtained from the IEEE by sending a request to pubs-permissions@ieee.org.}
\thanks{This work has been supported in part by the National Nature Science Foundation of China under Grant 62201347; and in part by Shanghai Sailing Program under Grant 22YF1428400. (\textit{Corresponding author: Yijie Mao})} 
\thanks{Y. Wang and Y. Mao are with the School of Information Science and Technology, ShanghaiTech University, Shanghai 201210, China (email: \{wangyw22022, maoyj\}@shanghaitech.edu.cn).}
\thanks{S. Ji is with the Department of Computer Science, The University of Hong Kong, Hong Kong 999077, China (email: sijieji@hku.hk).}}

\maketitle
\thispagestyle{empty}
\pagestyle{empty}

\begin{abstract}
Rate splitting multiple access (RSMA) relies on beamforming design for attaining spectral efficiency and energy efficiency gains over traditional multiple access schemes. While conventional optimization approaches such as weighted minimum mean square error (WMMSE) achieve suboptimal solutions for RSMA beamforming optimization, they are computationally demanding. A novel approach based on fractional programming (FP) has unveiled the optimal beamforming structure (OBS) for RSMA. This method, combined with a hyperplane fixed point iteration (HFPI) approach, named FP-HFPI, provides suboptimal beamforming solutions with identical sum rate performance but much lower computational complexity compared to WMMSE. Inspired by such an approach, in this work, a novel deep unfolding framework based on FP-HFPI, named rate-splitting-beamforming neural network (RS-BNN), is proposed to unfold the FP-HFPI algorithm. Numerical results indicate that the proposed RS-BNN attains a level of performance closely matching that of WMMSE and FP-HFPI, while dramatically reducing the computational complexity. 
\end{abstract}



\section{Introduction}

\par The advent of rate-splitting multiple access (RSMA) has been acknowledged as a burgeoning paradigm within the domain of multiple access (MA) for the sixth-generation mobile communications (6G) system. In contrast to established MA schemes, such as space division MA (SDMA) and non-orthogonal MA (NOMA), RSMA 
offers superior flexibility in interference management \cite{Mao2017RatesplittingMA}, and therefore excels over traditional MA methods in terms of spectral efficiency, energy efficiency, user fairness, and transmission reliability \cite{Mao2022Survey,10038476}. 
In order to attain all the aforementioned performance gains, RSMA requires a delicate beamforming design. The state-of-the-art beamforming design algorithms of RSMA are generally classified into two categories, namely, model-based and deep learning (DL)-based approaches. 
\par One of the most classic model-based beamforming algorithms for RSMA is the weighted minimum mean-square error (WMMSE) algorithm, which relies on the underlying mathematical transmission models as well as the estimation of the model parameters\cite{2021arXiv210104726S}. 
Recently, a novel optimization algorithm, named fractional programming--hyperplane fixed point iteration (FP-HFPI), has been proposed in \cite{fang2023optimal}. Such an approach first discovers the optimal beamforming structure (OBS) of the 1-layer rate-splitting (RS) based on fractional programming (FP) and then utilizes the hyperplane fixed point iteration (HFPI) approach to find the optimal Lagrangian dual variables. FP-HFPI is shown to achieve significant computational complexity reduction compared with WMMSE since it does not rely on any optimization toolbox such as CVX\cite{cvx}.
\par Another line of research for the beamforming design of RSMA is the DL-based algorithms, mainly considering end-to-end learning and deep unfolding as two major approaches. 
In \cite{Camana2022DeepLP}, an end-to-end deep neural network (DNN) is employed to find the common rate allocation of RSMA.
Compared to end-to-end learning, deep unfolding bridges the model-based methods and the pure data-driven methods, requiring smaller datasets while providing better interpretability. 
In \cite{Wu2022DeepLR} and \cite{Wang2022BatchGD}, two different deep unfolding algorithms based on the WMMSE algorithm are introduced for RSMA beamforming design. These algorithms aim to directly output either the beamforming vectors or their gradients, which however are difficult to predict due to their substantial dimensionality.
Domain knowledge, such as the spatial correlation \cite{doi:10.34133/space.0109} and OBS \cite{fang2023optimal,2023arXiv231216559F}, is therefore important in reducing the output dimension of the DL-based method.
To better bridge the model-based and DL-based approaches based on the domain knowledge, a novel beamforming neural network (BNN) is proposed in \cite{Xia2019ADL}. 
This approach utilizes the OBS for the SDMA weighted sum rate (WSR) maximization problem, requiring only a small-scale DNN for predicting the power allocation and Lagrangian dual variables. 
{However, this BNN is specifically tailored for SDMA-based systems, and it cannot be extended to address the WSR maximization problem of RSMA. This limitation arises from the inclusion of extra optimization variables for the beamforming vector of the common stream, along with extra Lagrangian dual variables related to the worst-case rate of the common stream.}
\par In this work, we aim to bridge the aforementioned research gaps and explore efficient DL-based beamforming approaches for RSMA. The main contributions are listed as follows:
\begin{itemize}
    \item We introduce a pioneering deep unfolding framework named rate-splitting-BNN (RS-BNN) to solve the RSMA sum rate (SR) maximization problem, specifically focusing on the beamforming design in multi-user multiple-input multi-user single-output (MU-MISO) networks.  {Unlike traditional pure data-driven approaches,  RS-BNN is based on the OBS of RSMA, successfully reducing the output dimension and simplifying the neural network design. To the best of our knowledge, this is the first work that explores the efficient DL-based beamforming approaches for RSMA that leverage its OBS to reduce the output dimension.}
    \item  {To marry the advantages of both model-based and data-driven approaches, the proposed RS-BNN identifies the optimal Lagrangian dual variables within the OBS of RSMA via unfolding the FP-HFPI algorithm and substituting the HFPI algorithm with a DNN featuring only one hidden layer.} In contrast to the HFPI approach, the proposed DL-based method achieves significantly reduced computational complexity without compromising SR performance. 
    \item  Numerical results show that the proposed RS-BNN maintains nearly identical SR performance as the state-of-the-art model-based FP-HFPI and WMMSE algorithms while significantly reducing the computational time. It also shows significant SR enhancement compared to the pure data-driven CNN-based baseline scheme.
\end{itemize}

\section{System Model of 1-layer RS}
\label{sec: system model}
We consider a downlink MU-MISO transmission network, in which a base station (BS) with $N_t$ transmit antennas delivers messages $\{W_1,...,W_K\}$ to $K$ single-antenna users indexed by $\mathcal{K}=\{1,...,K\}$. The most practical and simplest RSMA scheme, namely,  1-layer RS \cite{Mao2022Survey} is considered.\footnote{ {Although only the simplest 1-layer RS is considered in this paper, it should be noted that the approach proposed in this work can be directly extended to other complex RSMA frameworks such as 2-layer hierarchical RS and generalized RS \cite{Mao2022Survey}.}} This scheme involves the splitting of each message $W_k$ for user-$k$ into a common submessage $W_{0,k}$ and a private submessage $W_{p,k}$. The common message $W_0$, formed by combining all common submessages, is encoded into the common stream $s_0$, while the private submessages are respectively encoded into the private streams $\{s_1,...,s_K\}$. All transmitted streams, denoted as $\mathbf{s}=\left[s_0,s_1,...,s_K\right]^T\in\mathbb{C}^{(K+1)\times 1}$, where $\mathbb{E}\{\mathbf{s}\mathbf{s}^H\}=\mathbf{I}$, are preprocessed by the beamforming matrix $\mathbf{P}=\left[\mathbf{p}_0,\mathbf{p}_1,...,\mathbf{p}_K\right]\in\mathbb{C}^{N_t\times (K+1)}$ and transmitted to the users. 
Let $P_t$ denote the sum transmit power constraint such that the beamforming matrix satisfies $\text{tr}(\mathbf{P}\mathbf{P}^H)\leqslant P_t$, the transmit signal at the BS is
\begin{equation}
\mathbf{x}=\mathbf{p}_0s_0+\sum_{k\in\mathcal{K}}\mathbf{p}_ks_k.
 \end{equation}

For a given channel state information (CSI) $\mathbf{h}_k$, the received signal $y_k$ at user-$k$ is ${y}_k=\mathbf{h}_k^H\mathbf{x}+n_k$,
where $n_k\sim \mathcal{CN}(0,1)$ is the corresponding additive white Gaussian noise (AWGN) component. At user-$k$, the common stream $s_0$ is first decoded to attain the contained common submessage $\widehat{W}_{0,k}$. After removing the common stream $s_0$ from the received signal, user-$k$ then decodes the intended private stream $s_k$ to attain the private submessages $\widehat{W}_{p,k}$.
The signal-to-interference-noise ratios (SINRs) of decoding $s_0$ and $s_k$ at user-$k$ are 
   \begin{equation}
\begin{aligned}
\gamma_{0,k}=\frac{{\left|\mathbf{h}_k^H\mathbf{p}_0\right|}^2}{\sum_{i\in\mathcal{K}}{\left|\mathbf{h}_k^H\mathbf{p}_i\right|}^2+1},\gamma_{k}=\frac{{\left|\mathbf{h}_k^H\mathbf{p}_k\right|}^2}{\sum_{i\in\mathcal{K},i\neq k}{\left|\mathbf{h}_k^H\mathbf{p}_i\right|}^2+1}.
\end{aligned}
 \end{equation}With the SINRs, the decoding rates of $s_0$ and $s_k$ at user-$k$ are given as $R_{0,k}=\log_2(1+\gamma_{0,k})$ and $R_k=\log_2(1+\gamma_k)$, while the maximum achievable rate of $s_0$ is $R_0=\min_{k\in\mathcal{K}} \{R_{0,k}\}$. The SR expression is 
\begin{equation}
\label{eq:SR}
    SR(\mathbf{P})=R_0+\sum_{k\in\mathcal{K}}R_k.
\end{equation}
Therefore, the SR maximization problem is formulated as
\begin{equation}
\begin{aligned}
\label{prob:org}
\max_{\text{tr}(\mathbf{P}\mathbf{P}^H)\leqslant P_t}SR(\mathbf{P}).
 \end{aligned}
 \end{equation}

\section{Optimal Beamforming Structure of RSMA and the FP-HFPI algorithm}
\label{sec:FPHFPI}
In this section, the OBS of RSMA is first characterized followed by a detailed illustration of the FP-HFPI algorithm introduced in \cite{fang2023optimal}. This algorithm is instrumental in computing the optimal dual variables by leveraging the established OBS.

\textit{Optimal beamforming structure of RSMA:} To obtain the  OBS of RSMA, the original SR maximization problem (\ref{prob:org}) is first transformed into a block-wise convex problem through the FP approach. Then, the optimal beamformer is obtained by the Karush-Kuhn-Tucker (KKT) conditions of the transformed problem. 
\par
The FP framework, as introduced in \cite{Shen2018FractionalPF}, effectively addresses the nonconvexity resulting from the fractional SINR expressions. By utilizing the Lagrangian dual transform and quadratic transform, the expressions of $R_{0,k}$ and $R_k$ are transformed into 
     \begin{equation}
\begin{aligned}
g_{i,k}(\mathbf{P},\alpha_{i,k},\beta_{i,k})&= \log_2(1+\alpha_{i,k})-\alpha_{i,k}\\&+2\sqrt{1+\alpha_{i,k}}\Re\left\{\beta_{i,k}^H\mathbf{h}_k^H\mathbf{p}_k\right\}\\&-{\left| \beta_{i,k}\right|}^2\left(\sum_{j\in\{i\}\cup\mathcal{K}}{\left|\mathbf{h}_k^H\mathbf{p}_j\right|}^2+\sigma_k^2\right),
\end{aligned}
 \end{equation}where $\alpha_{i,k}$ is the auxiliary variable corresponding to the SINR $\gamma_{i,k}$ and $\beta_{i,k}$ is the auxiliary variable introduced to decouple the fractional term. As the private stream is only decoded by the corresponding user, the notations $\alpha_{k,k}$, $\beta_{k,k}$, $g_{k,k}$, and $\gamma_{k,k}$ are further simplified as $\alpha_k$, $\beta_k$, $g_k$, and $\gamma_{k}$. Let $\mathcal{S}_{\mathbf{P}}=\{\mathbf{P} \mid \text{tr}(\mathbf{P}\mathbf{P}^H)\leqslant P_t\}$, $\mathcal{S}_\alpha=\{\alpha_{0,k},\alpha_k\mid\forall k\in\mathcal{K}\}$ and $\mathcal{S}_\beta=\{\beta_{0,k},\beta_k\mid\forall k\in\mathcal{K}\}$, the original SR maximization problem (\ref{prob:org}) is equivalently transformed into 
 \begin{equation}
\begin{aligned}
\label{problem_g}
\max_{\mathcal{S}_{\mathbf{P}},\mathcal{S}_\alpha,\mathcal{S}_\beta} (\min_{k\in\mathcal{K}}g_{0,k}+\sum_{k\in\mathcal{K}}g_k)
\end{aligned}
 \end{equation}
Problem (\ref{problem_g}) is block-wise convex. The optimal $\alpha_{i,k}^\star$ and $\beta_{i,k}^\star$ are respectively derived by fixing other two blocks of variables \cite{fang2023optimal}, which are given as 
 \begin{equation}
 \begin{aligned}    
 \label{eq:beta}
 \alpha_{i,k}^\star=\gamma_{i,k},\text{  }\beta_{i,k}^\star=\frac{\sqrt{1+\alpha_{i,k}^\star}\mathbf{h}_k^H\mathbf{p}_i}{\sum_{j\in\{i\}\cup\mathcal{K}}{|\mathbf{h}_k^H\mathbf{p}_j|}^2+\sigma_k^2}.
 \end{aligned}
\end{equation}
By fixing variables in $\mathcal{S}_\alpha$ and $\mathcal{S}_\beta$, and replacing the term $\min_{k\in\mathcal{K}}g_{0,k}$ with a slack variable $y$,  problem (\ref{problem_g}) is then simplified as
\begin{subequations}
\label{eq:convex}
	\begin{align}
		\max_{\mathbf{P},y} \,\,\,&y+\sum_{k\in\mathcal{K}} {g}_k(\mathbf{P}),  \label{convex_obj} \\
	\mbox{s.t.}\,\,
	&\,\,	\text{tr}(\mathbf{P}\mathbf{P}^{H})\leqslant P_{t}, \label{SAAconvex_P}\\
	&  \,\,\ y\leqslant {g}_{0,k}(\mathbf{P}), \forall k\in\mathcal{K}.\label{convex_y}
\end{align}
\end{subequations}
Problem (\ref{eq:convex}) is concave. Denote $\mu$ and $\bm{\lambda}=\left[\lambda_1,...,\lambda_K\right]$ as the Lagrangian dual variables for the corresponding constraints (\ref{SAAconvex_P}) and (\ref{convex_y}), the Lagrangian function for problem (\ref{eq:convex}) is
\begin{equation}
    \begin{aligned}
L(\mathbf{P},\bm{\lambda},y,\mu)&= \sum_{k\in\mathcal{K}}g_k(\mathbf{P})+y-\sum_{k\in\mathcal{K}}\lambda_k(y-g_{0,k}(\mathbf{P}))\\
&-\mu(\text{tr}(\mathbf{P}^H\mathbf{P})-P_t).
    \end{aligned}
\end{equation}
As problem (\ref{eq:convex}) is concave and also strictly feasible, the optimal solution exists and satisfies the Karush-Kuhn-Tucker (KKT) conditions.  {Therefore, we establish the following Theorem 1, which shows the optimal solution structure $\mathbf{P}^\star$ for both problems (\ref{eq:convex}) and (\ref{prob:org}).}

\textbf{Theorem 1.} \textit{The OBS for the subproblem \eqref{eq:convex} and the SR maximization problem (\ref{prob:org}) of 1-layer RS is given as \cite{fang2023optimal}}
\begin{equation}
    \begin{aligned}\label{eq:opt_precoders}
    \,\resizebox{0.46\textwidth}{!}{$\mathbf{p}_0^\star=\,{\left(\sum_{j\in\mathcal{K}}\lambda_j^\star{|\beta_{0,j}|}^2\mathbf{h}_j\mathbf{h}_j^H+\mu^\star \mathbf{I}\right)^{-1}}\sum_{j\in\mathcal{K}} \sqrt{1+\alpha_{0,j}}\beta_{0,j}\lambda_j^\star\mathbf{h}_j$},\\
    \resizebox{0.46\textwidth}{!}{$\mathbf{p}_k^\star={\left(\sum_{j\in\mathcal{K}}({|\beta_j|}^2+\lambda_j^\star{|\beta_{0,j}|}^2)\mathbf{h}_j\mathbf{h}_j^H+\mu^\star\mathbf{I}\right)^{-1}}\sqrt{1+\alpha_k}\beta_k\mathbf{h}_k$},\\
    \end{aligned}
\end{equation}
\textit{where $\bm{\lambda}^\star$ and $\mu^\star$ are the optimal dual variables.}
\begin{figure}
\centering
\includegraphics[width=0.43\textwidth]{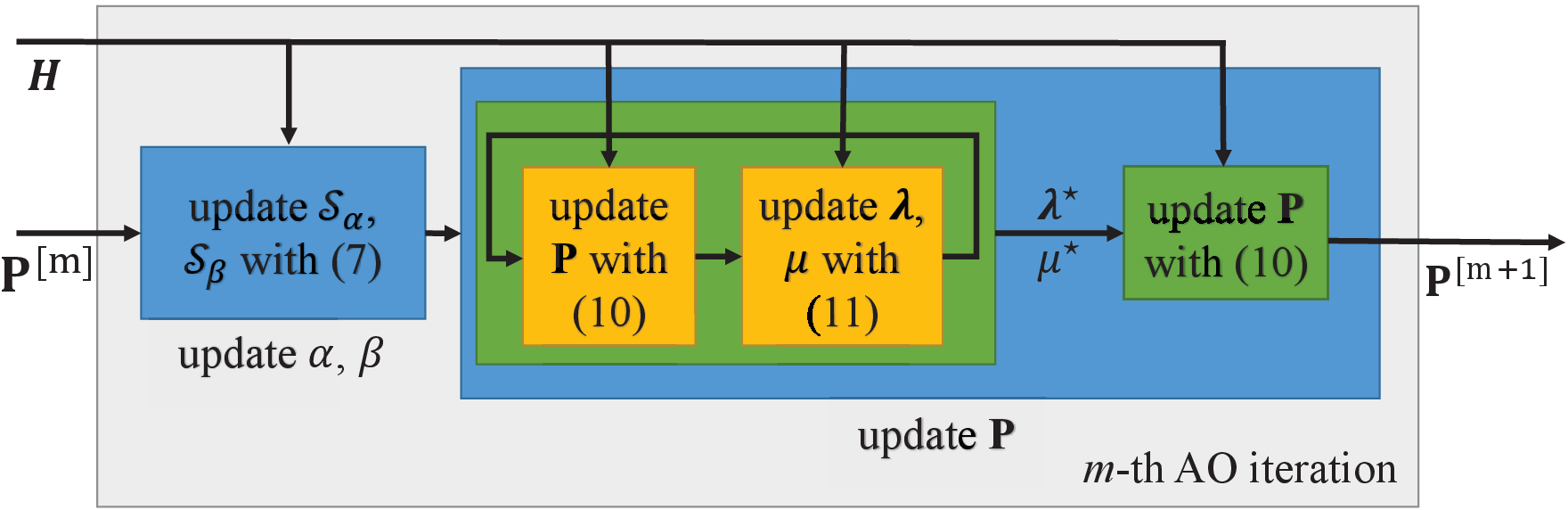}
\caption{The procedure of each AO iteration for the FP-HFPI algorithm.}
\label{fig:FPHFPI}
\end{figure}

\textit{Proof:}
 Following the proof in \cite{fang2023optimal}, the optimal solution  $\mathbf{P}^\star$ is guaranteed to exist and is shared between the convex subproblem (\ref{eq:convex}) and the original SR maximization problem (\ref{prob:org}). According to the KKT conditions of (\ref{eq:convex}), the optimal $\mathbf{P}^\star$ of problem \eqref{eq:convex} can be derived from $\frac{\partial L}{\partial y}=0$ and $\frac{\partial L}{\partial \mathbf{p}_k}=\mathbf{0}$. By solving $\frac{\partial L}{\partial \mathbf{p}_k}=\mathbf{0}$ with the requirement $\sum_{k\in\mathcal{K}}\lambda_k=1$ drawn from $\frac{\partial L}{\partial y}=0$, we then obtain the optimal $\mathbf{p}_k^\star$ and $\mathbf{p}_0^\star$ in \eqref{eq:opt_precoders}.  {Moreover, with the optimal $\mathbf{P}^\star$, the optimal $y^{\star}$ for problem (8) can be directly obtained by $y^\star=\min_{k\in\mathcal{K}}g_{0,k}(\mathbf{P^\star})$.} Due to space constraints, the detailed calculation procedure is not specified here, interested readers are referred to \cite{fang2023optimal} for more comprehensive explanations. $\hfill\blacksquare$



\begin{figure*}
\centering
\includegraphics[width=0.73\textwidth]{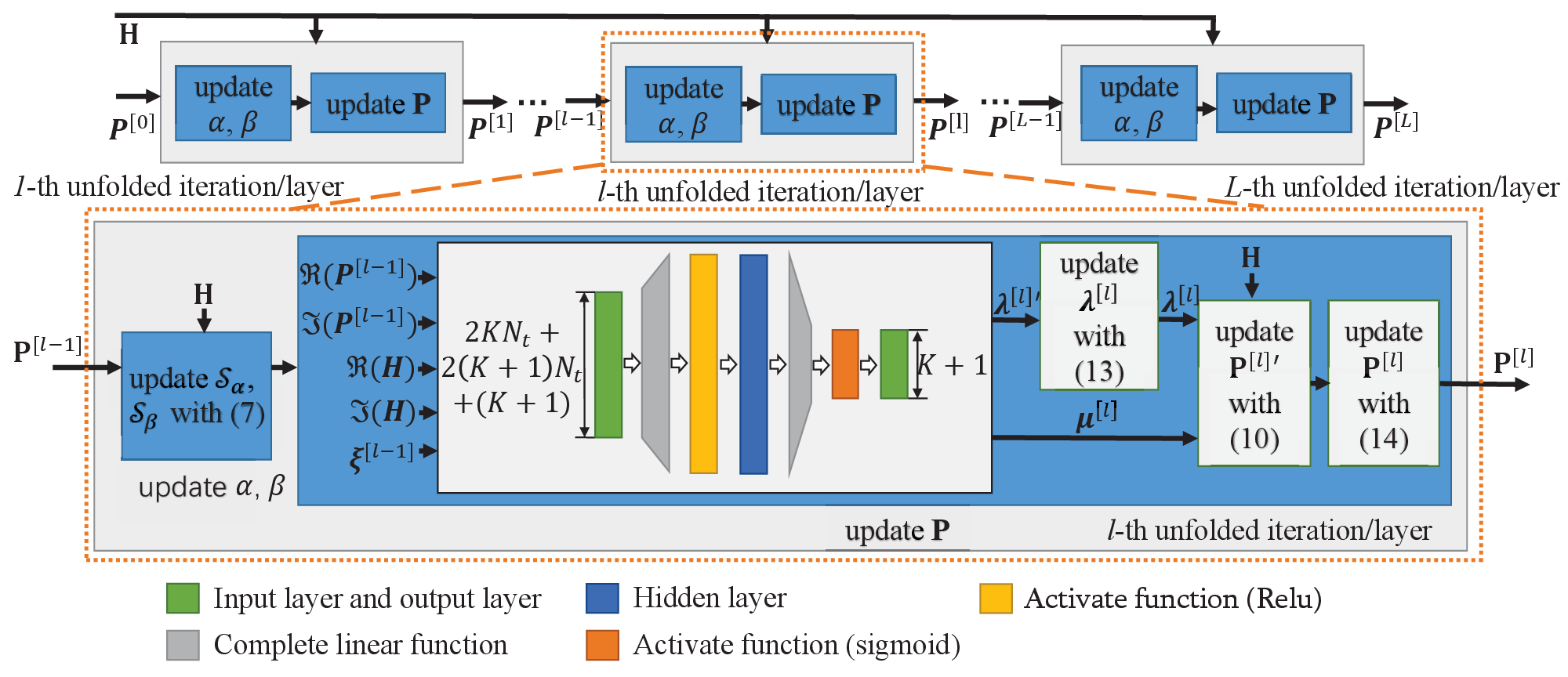}
\caption{The proposed RS-BNN framework.}
\label{fig:overall}
\end{figure*}

\par \textit{Hyperplane fixed point iteration algorithm:} As shown in equation (\ref{eq:opt_precoders}), the optimal solutions of $\mathbf{P}^\star$ and $y^\star$ rely on the optimal dual variables $\bm{\lambda}^\star$ and $\mu^\star$, which however, are rather challenging to obtain. In \cite{fang2023optimal}, an HFPI method is therefore proposed to find the optimal dual variables. This iterative algorithm is designed based on fixed point iteration. The procedure is summarized as follows. 
\par Let $t$ represent the iteration time in the HFPI process, and $\mathbf{P}^{[t]}$, $\bm{\lambda}^{[t]}$, $\mu^{[t]}$ as the value of $\mathbf{P}$, $\bm{\lambda}$, and $\mu$ at iteration $t$. The value of $g_{0,k}(\mathbf{P}^{[t]})$ is simply denoted as $h_{0,k}^{[t]}$. Denote $\widehat{k}$ as the user index with the worst-case $h_{0,k}^{[t]}$, i.e., $h_{0,\widehat{k}}^{[t]}=\min_{k\in\mathcal{K}}h_{0,k}^{[t]}$. Following \cite{fang2023optimal}, by introducing $\rho$ as the non-negative constant to ensure the convergence, the Lagrangian dual variables are updated based on HFPI iteratively, which are given as
\begin{equation}
\begin{aligned}
\label{equ:opt_lagrange}
\lambda_{k}^{[t+1]} & =\frac{h_{0, \widehat{k}}^{[t]}+\rho}{h_{0, k}^{[t]}+\rho} \lambda_{k}^{[t]}, \quad k \neq \widehat{k}, \forall k \in \mathcal{K}, \\
\lambda_{\widehat{k}}^{[t+1]} & =\lambda_{\widehat{k}}^{[t]}+\sum_{k\in\mathcal{K}}\left(1-\frac{h_{0, \widehat{k}}^{[t]}+\rho}{h_{0, k}^{[t]}+\rho}\right) \lambda_{k}^{[t]}, \\
\mu^{[t+1]} & =\frac{\operatorname{tr}\left((\mathbf{P}^{[t]})^{H} \mathbf{P}^{[t]}\right)+\rho}{P_{t}+\rho} \mu^{[t]}.
\end{aligned}
\end{equation}
Based on the OBS of RSMA and the aforementioned HFPI algorithm to update the dual variables, we obtain the FP-HFPI algorithm. In general, there are two loops in the FP-HFPI algorithm: an outer loop and an inner loop. One iteration of the outer loop which contains an inner loop is considered as one alternative optimization (AO) iteration $m$. As shown in Fig. \ref{fig:FPHFPI}, one AO iteration  includes the update of  $\mathcal{S}_\alpha$,  $\mathcal{S}_\beta$, and $\mathbf{P}$. Moreover, the inner loop in each AO iteration $m$ is used to obtain the optimal $\bm{\lambda}^\star$ and $\mu^\star$ for the current iteration. With the optimal $\bm{\lambda}^\star$ and $\mu^\star$, $\mathbf{P}^{[m+1]}$  is then updated based on the OBS in (\ref{eq:opt_precoders}). After repeating the AO iterations, FP-HFPI finally converges to the sub-optimal solution of problem (\ref{prob:org}).

\section{Proposed Deep Unfolding Framework}
\label{sec:DeepUnfolding}

\par 
This section introduces the RS-BNN DL framework and provides a detailed discussion of its training process.
\subsection{The Proposed RS-BNN Framework}
\par 
 In the FP-HFPI algorithm,  the inner loop within each AO iteration takes considerable computational time. Therefore, in the proposed framework, we replace the inner loop of FP-HFPI with a small-scale DNN to predict $\bm{\lambda}^\star$ and $\mu^\star$, and unfold the outer loop of FP-HFPI with a fixed number of unfolded iterations/layers. 

\par We use $\mathcal{F}_l$ to denote the DNN for the $l$-th unfolded iteration/layer. The map function of $\mathcal{F}_l$ can be written as 
\begin{equation}
\label{eq:RSBNN}
    \begin{aligned}
        \left[{\bm{\lambda}}^{[l]'},{\mu}^{[l]}\right]=\mathcal{F}_l\left(\mathbf{H},{\mathbf{P}}^{[l-1]},{\bm{\lambda}}^{[l-1]},{\mu}^{[l-1]}\right),
    \end{aligned}
\end{equation}where the ${\bm{\lambda}}^{[l]'}$ is the preliminary value of ${\bm{\lambda}}^{[l]}$. To ensure that $\bm{\lambda}$ satisfies the KKT condition $\sum_{k\in\mathcal{K}}\lambda_k=1$ of problem (\ref{eq:convex}), ${\bm{\lambda}}^{[l]'}$ is further adjusted via
\begin{equation}
\label{eq:RSBNN_lambda}
    \begin{aligned}
{\bm{\lambda}}^{[l]}=\frac{{\bm{\lambda}}^{[l]'}+\epsilon}{\sum_{k\in\mathcal{K}}{\lambda}_k'+k\epsilon},
    \end{aligned}
\end{equation}
where $\epsilon$ is a small constant to guarantee ${\lambda}_k>0$. This is also required to satisfy the KKT conditions.
Based on (\ref{eq:RSBNN}) and (\ref{eq:RSBNN_lambda}), we obtain the Lagrangian dual variable vector ${\bm{\xi}}^{[l]}=[{\bm{\lambda}}^{[l]},{\mu}^{[l]}]$, with which beamformer ${\mathbf{P}}^{[l]'}$ can be calculated by equation (\ref{eq:opt_precoders}) as the preliminary predicted value of ${\mathbf{P}}^{[l]}$. Note that the prediction of the DNN may introduce bias, leading to ${\mathbf{P}}^{[l]'}$ not adhering to the power constraint. To ensure the power constraint \eqref{SAAconvex_P} holds, the output of the $l$-th unfolded iteration, i.e., ${\mathbf{P}}^{[l]}$, is rectified as
\begin{equation}
\label{Normal_P}
    {\mathbf{P}}^{[l]}=\sqrt{\frac{{P_t}}{\text{tr}\left({\mathbf{P}}^{[l]'}{{\left({\mathbf{P}}^{[l]'}\right)}}^H\right)}}{\mathbf{P}}^{[l]'}.
\end{equation}

\par Fig. \ref{fig:overall} illustrates the proposed deep-unfolding framework. Specifically, we unfold $L$ AO iterations of the FP-HFPI algorithm, substituting each AO iteration $l$ 
with a layer-wise structure that takes $\mathbf{P}^{[l-1]}$ as input and produces $\mathbf{P}^{[l]}$ as output with the help of a simple DNN.
According to equation (\ref{equ:opt_lagrange}), the optimal Language dual variables ${\bm{\xi}}^{[l]}=[{\bm{\lambda}}^{[l]},{\mu}^{[l]}]$ of the $l$-th unfolded iteration/layer rely on the CSI $\mathbf{H}$, the Lagrangian dual variable vector ${\bm{\xi}}^{[l-1]}$, and the beamformers ${\mathbf{P}}^{[l-1]}$ obtained from the $l-1$-th unfolded iteration/layer. As both $\mathbf{H}$ and ${\mathbf{P}}^{[l-1]}$ are complex matrices that are not suitable for the DNN to process, we separate their real parts and imaginary parts as $\Re(\mathbf{H})$, $\mathfrak{I}(\mathbf{H})$, $\Re({\mathbf{P}}^{[l-1]})$ and $\mathfrak{I}({\mathbf{P}}^{[l-1]})$.
As illustrated in Fig. \ref{fig:overall}, the DNN for the $l$-th unfolded iteration consists of one input layer, one hidden layer, and one output layer.  {The hidden layer contains 512 neurons.} Two types of activation functions, namely, Relu and sigmoid, are deployed. The input layer receives the channel matrix $\mathbf{H}$, the beamformer ${\mathbf{P}}^{[l-1]}$ and the combined Lagrangian dual variables vector ${\bm{\xi}}^{[l-1]}$ with a input dimension of $2KN_t+2(K+1)N_t+(K+1)$, while the output layer generates a vector of length $K+1$. The input and output layers are respectively connected with the hidden layer by a complete linear function.


\subsection{Model Training}
The training process of the proposed RS-BNN is discussed in this subsection. 
As the HFPI algorithm cannot guarantee convergence to the globally optimal Lagrangian dual variables, depending solely on FP-HFPI training data may result in locally optimal solutions.
To overcome such limitation, we follow \cite{Xia2019ADL} and train the framework by combining supervised and unsupervised learning to achieve better performance. 
Specifically, we first train based on supervised learning with labels obtained from the FP-HFPI algorithm. This yields solutions approximating those of the FP-HFPI algorithm. To further boost performance, we then train RS-BNN by unsupervised learning, employing a loss function based on the SR metric.
\subsubsection{Supervised Learning}
As outlined in the proposed RS-BNN framework, the prediction of Lagrangian dual variables involves $L$ unfolded iterations/layers of DNNs. Unlike FP-HFPI, which adjusts the number of iterations in the FP-HFPI algorithm according to the tolerance of convergence, RS-BNN maintains a fixed number of unfolded iterations. As the solutions of the FP-HFPI are sub-optimal, over-reliance on them may lead the DNN in the wrong direction. Therefore, we only use the values of the Lagrangian dual variables in the first and the last iterations of HFPI  as labels to guide the training of RS-BNN. 
To optimize trainable parameters in RS-BNN, the loss function during the supervised learning phase is defined as

\begin{equation}
    {Loss}_{supervised}={\|{\bm{\xi}}^{[0]}-\bm{\xi}_{first}\|}^2+{\|{\bm{\xi}}^{[L]}-\bm{\xi}_{last}\|}^2
\end{equation}
where $\bm{\xi}_{first}$ and $\bm{\xi}_{last}$ are labels obtained by the FP-HFPI algorithm, ${\bm{\xi}}^{[0]}$ and ${\bm{\xi}}^{[L]}$ are the predicted results of the first and the last unfolded iterations of the proposed RS-BNN framework, respectively. We don't provide guidelines for the outputs of the unfolded iterations in between, in order to give the DNN more freedom to explore better solutions.
\subsubsection{Unsupervised Learning}
As mentioned earlier, to address the locally optimal solutions obtained by FP-HFPI, RS-BNN undergoes unsupervised learning-based training after the supervised learning-based training. The corresponding loss function takes the SR expression directly as 
 {
\begin{equation}
    {Loss}_{unsupervised}=-SR({\mathbf{P}}^{[L]}),
\end{equation}}
where ${\mathbf{P}}^{[L]}$ is the output of the \textit{L}-th unfolded iteration.

\section{Numerical Results}
In this section, we evaluate the performance of the proposed RS-BNN framework. 
The comparisons include the following four schemes:
\begin{itemize}
\item \textbf{RS-BNN}---This is the deep unfolding framework proposed in Section \ref{sec:DeepUnfolding}.  {By utilizing the OBS, each iteration/layer contains only a DNN with one hidden layer to predict the Lagrangian variables.  The output dimension of the neural network is $K+1$, which does not scale with the number of transmit antennas $N_t$. }
\item \textbf{WMMSE}---This is a model-based algorithm that has been widely applied to design beamforming for RSMA-aided networks \cite{Mao2017RatesplittingMA}. It requires an optimization toolbox, i.e., CVX \cite{cvx}, to solve convex subproblems iteratively.
\item \textbf{FP-HFPI}---This is the model-based algorithm as specified in Section \ref{sec:FPHFPI} \cite{fang2023optimal}. It does not require any optimization toolbox but requires two loops to obtain suboptimal solutions to problem \eqref{prob:org}. 
\item \textbf{Black-box CNN}---Inspired by \cite{Huang2020FastBeamforming}, we introduce another baseline scheme, which is a pure data-driven algorithm based on the black-box convolutional neural networks (CNN) to solve the SR maximization problem of RSMA. The CNN takes as input the channel matrix $\mathbf{H}$ rearranged as $[\Re(\mathbf{H}), \Im(\mathbf{H})]$, and produces an output of $[\Re(\mathbf{P}), \Im(\mathbf{P})]$. This allows us to make predictions about the complex beamforming matrix $\mathbf{P}$. During the supervised learning stage, we utilize the mean-square error (MSE) loss function, while in the unsupervised learning stage, we use the SR loss function.  {The output dimension of such black-box CNN is $2N_t(K+1)$, which is much larger than the proposed RS-BNN.}
\end{itemize}
 {The computational complexity comparison between RS-BNN, WMMSE, FP-HFPI, and black-box CNN is shown in Table I. $L$ is the number of unfolded iterations for RS-BNN and $M$ represents the number of neurons in the hidden layer within each unfolded iteration of RS-BNN. The number of outer and inner iterations for WMMSE and FP-HFPI are respectively denoted as $I_{outer}$ and $I_{inner}$. The number of the convolutional layers for the black-box CNN is denoted as ${L^{'}}$, the size of the convolution kernel is denoted as $N_k$, and the number of input and output channels is respectively denoted as $C_{in}$ and $C_{out}$.} 

\begin{table}[htbp]
\caption{ {The Quantified Computational Complexities of All Schemes.}}
\centering
    \begin{tabular}{|c|c|}
\hline
 {Schemes}       &  {Computational complexity} \\ \hline
 {RS-BNN }       &  {$\mathcal{O}(LMKN_t+LN_t^2K+LK^2N_t)$ }              \\ \hline
 {WMMSE}         &  {$\mathcal{O}(I_{outer}I_{inner}{[KN_t]}^{3.5})$} \\ \hline
 {FP-HFPI}       &  {\makecell[c]{$\mathcal{O}(I_{outer}I_{inner}(N_t^3+KN_t^2+K^2N_t))$}}\\ \hline
 {Black-box CNN} &  {$\mathcal{O}({L^{'}}{(KN_t)}^2N_k^2C_{in}C_{out})$}                \\ \hline
\end{tabular}
\label{table:complexity}

\end{table}
\par Assume that the BS is fixed in the center of a cell of radius $100$ meters, while the users are randomly dropped within this cell. Let $\alpha$ represent the path-loss exponent, $d_k$ denote the distance between the base station and user $k$, and $d_0$ represent the reference distance. The channel of each user is modeled as $\mathbf{h}_k=\sqrt{\rho_k}\Tilde{\mathbf{h}}_k$, where $\rho_k=\frac{1}{1+{(\frac{d_k}{d_0})}^\alpha}$ is the large-scale path loss, and $\Tilde{\mathbf{h}}_k\sim\mathcal{CN}(\mathbf{0},\mathbf{I})$ is the small-scale fading.

\par The training data for both RS-BNN and the black-box CNN are produced by FP-HFPI with $\rho=0.1$, the inner loop tolerance $10^{-5}$, and the outer loop tolerance $10^{-4}$. The beamforming initialization of FP-HFPI and RS-BNN follows  \cite{fang2023optimal}. Due to the substantial computational time required by the WMMSE algorithm to generate a solution, obtaining results of WMMSE for a large test dataset is challenging. To ensure a fair comparison, we reduce the size of the test dataset. For our simulation, we generated 20,000 training samples and 100 testing samples, respectively. 
The validation set comprises 20$\%$ of the training set. All numerical results below are conducted using Windows 10 (64 b) operating system and Python 3.10. The training is implemented with Pytorch on a PC equipped with an NVIDIA A40 and the testing is processed on the CPU in order to guarantee a fair comparison.  {In our simulation, we have tried to mitigate the influence of the hardware and coding skills by running the simulation of all schemes on the same machine with the same version of Python and maintaining consistency in our code style. For example, since the proposed RS-BNN unfolds the FP-HPFI algorithm, we share the same codes for both RS-BNN and FP-HFPI to calculate the auxiliary variables and update the beamforming vectors $\mathbf{P}$ with the given Lagrangian variables. Moreover, both RS-BNN and the black-box CNN baseline employ an identical coding framework for the training and testing phases, utilizing the MSE and SR loss functions. The primary difference between RS-BNN and the black-box CNN lies in their neural network structure. Specifically, RS-BNN unfolds the FP-HFPI, while the black-box CNN solely utilizes a CNN.}
\par The architecture of the black-box CNN follows \cite{Xia2019ADL}, except that the output dimension is changed to $2N_t(K+1)$ and the activation function in the last layer of the CNN is replaced by equation (\ref{Normal_P}). RS-BNN is configured with $L=5$ unfolded iterations/layers, and $\epsilon$ in equation (\ref{eq:RSBNN_lambda}) is set to $0.01$. The dimension of the hidden layer is set to $512$.
We use the Adam optimizer for training. The batch size for training is fixed at $1000$, and the learning rate is $0.0001$. The training epoch is $200$, the supervised learning-based training phase takes $50$ epochs and the unsupervised phase takes another $150$ epochs. Additionally, an early stop scheme is implemented during training.  Specifically, if the validation loss fails to decrease for $7$ consecutive epochs, the training process ends.

\begin{figure}
\centering
\subfigure[SR versus $N_t$ and $K$.]{\label{fig:subfig:a}
\includegraphics[width=0.233\textwidth]{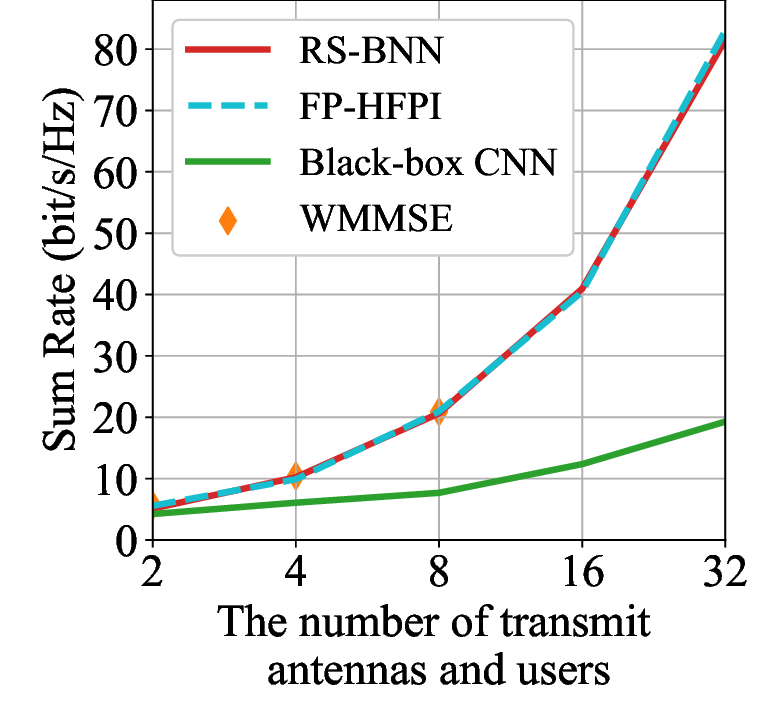}}
\subfigure[CPU Time versus $N_t$ and $K$.]{\label{fig:subfig:b}
\includegraphics[width=0.233\textwidth]{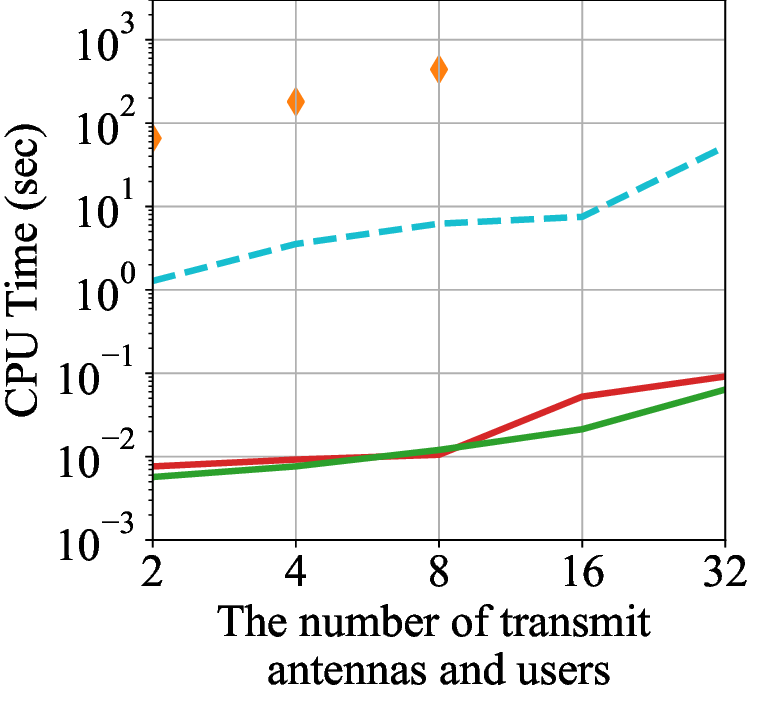}}
\caption{Sum rate or CPU time versus the number of transmit antennas and users ($N_t=K$) for different algorithms. SNR$=20$ dB.}
\label{fig: Sim1}
\end{figure}
\begin{figure}
\centering
\subfigure[SR versus SNR.]{\label{fig:subfig:c}
\includegraphics[width=0.23\textwidth]{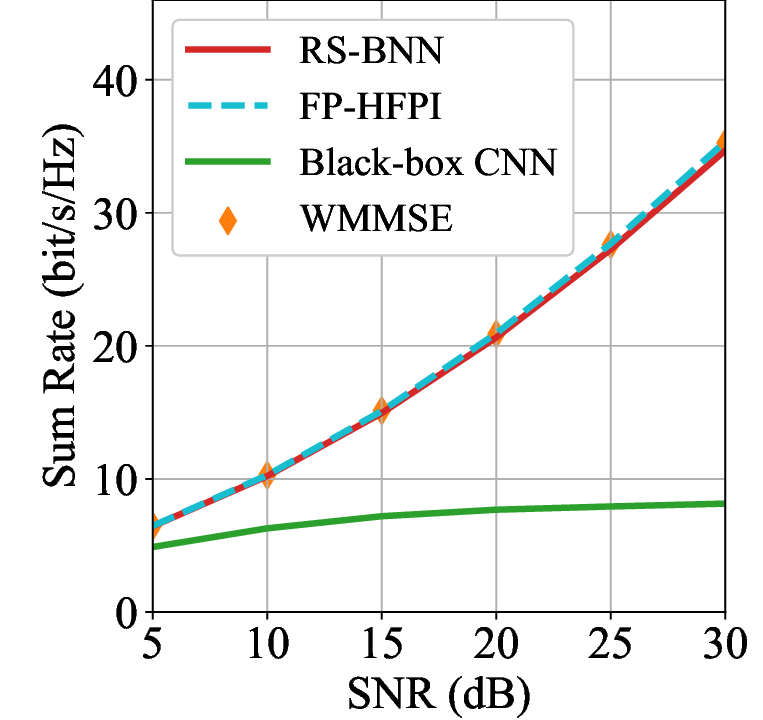}}
\hspace {0.5mm}
\subfigure[CPU time versus SNR.]{\label{fig:subfig:d}
\includegraphics[width=0.23\textwidth]{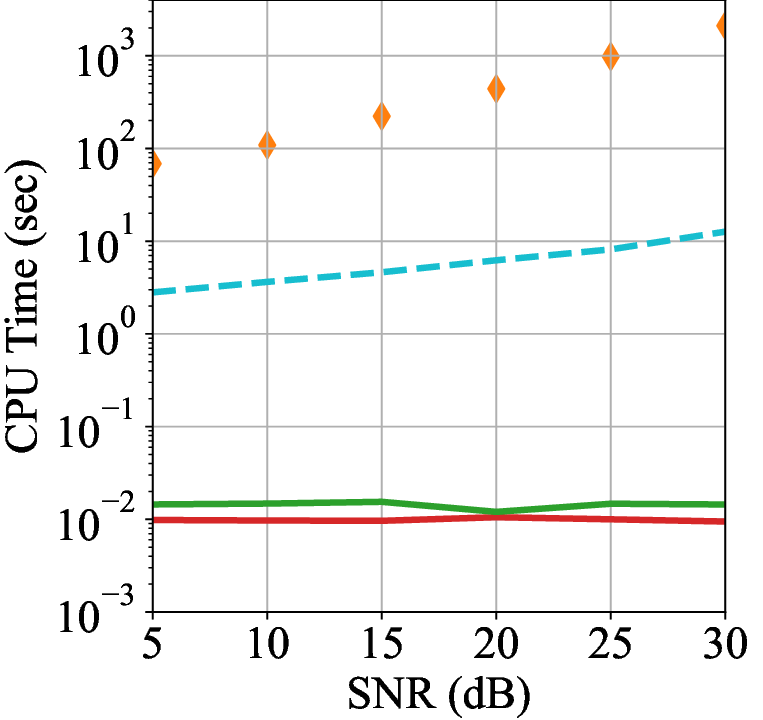}}
\caption{Sum rate or CPU time versus SNR comparison for different algorithms. $N_t=K=8$.}
\label{fig: Sim2}
\end{figure}

\par 
In Fig. \ref{fig: Sim1}, the system SR performance and CPU time are compared for all schemes when the number of transmit antennas and users are equal, i.e., $N_t=K$, and they span the range from $2$ to $32$ antennas/users. The transmit signal-to-noise (SNR), which is defined as $SNR=\frac{P_t}{\sigma_k^2}$ is set to $20$ dB. 
Due to the significant computational time demanded by the CVX solver, we present only partial baseline results of WMMSE when $N_t=K\leqslant8$.
It is shown in Fig. \ref{fig: Sim1} that our proposed RS-BNN provides nearly identical  SR performance as FP-HFPI and WMMSE, yet with a CPU time that is two orders of magnitude lower than that of FP-HFPI. This is due to the fact that RS-BNN only requires $L=5$ of unfolded iterations to obtain the near optimal beamforming solution while FP-HFPI requires an average of $197$ outer loop AO iterations and $23$ inner loop iterations within each outer loop AO iteration to obtain the solution. The black-box CNN shares similar time consumption as RS-BNN but it achieves much worse SR performance compared with the other three schemes, especially when there is a large number of users and transmit antennas.  With the help of the OBS of RSMA, RS-BNN effectively mitigates this issue.
\par In Fig. \ref{fig: Sim2}, we set $N_t=K=8$ and compare the SR performance and execution time when SNR increases from 5 dB to 30 dB. We observe from Fig. \ref{fig: Sim2} that the black-box CNN again fails to achieve satisfactory SR performance. In contrast, RS-BNN achieves nearly the same SR as FP-HFPI and WMMSE in all SNR regimes with a much lower CPU time than WMMSE and FP-HFPI. RS-BNN successfully finds the near optimal dual variables for the RSMA OBS, bypassing the computational burden associated with the two iterative loops in FP-HFPI or the use of the optimization toolbox in WMMSE. Therefore, the proposed RS-BNN brings notable advantages of decreasing computational time without compromising the SR performance.

\section{Conclusion}
In this study, a novel DL-based framework called RS-BNN is proposed for RSMA beamforming design by leveraging the OBS of RSMA. Our approach involves unfolding the established FP-HFPI algorithm and substituting HPFI with a small-scale DNN to derive the optimal Lagrangian dual variables. This unfolding process enables the proposed RS-BNN to efficiently design the near optimal beamformers for the SR maximization problem of 1-layer RS. Numerical results show the proposed RS-BNN significantly reduces the computational time while maintaining almost the same SR performance compared to WMMSE or FP-HFPI.  {The extension to the partial CSIT scenario and other RSMA frameworks will be considered as our future work.}

\bibliographystyle{IEEEtran}  
\bibliography{reference.bib}

\end{document}